\documentclass[a4paper]{article}

\usepackage{INTERSPEECH2019}
\usepackage[colorlinks,linkcolor=blue]{hyperref}
\usepackage{stfloats}

\title{Comprehensive Analysis and Exclusion Hypothesis of $\alpha$-Approximation Method for Discretizing Analog Systems}
\name{Shen Chen \textsuperscript{1, 2}, Jisong Wang\textsuperscript{2}, Dejun Liu\textsuperscript{2}, Jiaxi Ying\textsuperscript{2}, Shuai Wang\textsuperscript{2}}
\address{
    \begin{tabular}{c} 
       \textsuperscript{1}School of Electrical Engineering, Xi'an Jiaotong University \\ 
       \textsuperscript{2}SolaX Power Network Technology (Zhejiang) Co., Ltd.
    \end{tabular}
}
\email{chenshenpe@163.com}

\begin{document}

\maketitle

\begin{abstract}
A popular method for designing digital models is transforming the transfer function of the corresponding analog models 
from continuous domain (s-domain) into discrete domain (z-domain) using the s-to-z transformation.
The $\alpha$-approximation is a generalized form of these transformations.
When $\alpha$ is set to 0.5, the result is the well-known Tustin transformation or bi-linear transformation.
In this paper, we provided a comprehensive analysis of the $\alpha$-approximation method, 
including mathematical interpretation, stability analysis and distortion analysis.
Through mathematical interpretation, we revealed that it can be derived by numerically integrating the error function 
We defined this as the hexagonal approximation.
We demonstrated that the stable range of $\alpha$ was [0.5, 1] by doing stability analysis.
Through distortion analysis, we found that minimizing amplitude and phase distortion simultaneously seemed impossible 
by regulating $\alpha$ alone.
Finally, We proposed an exclusion hypothesis hypothesizing that there is no single parameter $\alpha$ to minimize the amplitude distortion and phase distortion 
simultaneously across all frequency points within the Nyquist frequency range.
This paper demonstrates that designing parameter $\alpha$ involves balancing amplitude and phase distortion.
\end{abstract}


\section{Introduction}
Nowadays, controllers and filters in modern digitally-controlled systems are typically designed in the continuous domain, 
and then transformed into the discrete domain using the S-to-Z transformation.
In machine learning field, applications increasingly rely on discretized models for training and inference. 
For instance, discretized continuous-time neural networks improve the stability and efficiency of deep learning models\textsuperscript{\cite{gu2024mamba}}.
Techniques such as neural ordinary differential equations (Neural ODEs) use bilinear discretization 
to preserve gradient flow during back-propagation\textsuperscript{\cite{mohammad25-DT}}.
However, all discretization methods introduce unwanted errors including amplitude and phase distortion.
Furthermore, some methods may even cause the discretized system to become unstable.
Therefore, discretization methods must be chosen carefully during the digital implementation process.

Between 2015 and 2024, research has focused on refining classical methods,
such as zero-order hold (ZOH)\textsuperscript{\cite{Ma18-ZOH}}, 
bilinear (Tustin) transformation\textsuperscript{\cite{Hoja08-Tustin},\cite{9256625},\cite{4089107}},  
and higher-order hold (HOH) techniques\textsuperscript{\cite{Zhang2014Discretization}},
while addressing emerging challenges related to nonlinear systems, digital twins, and real-time applications.
Zero-order hold continues to be the default method for digital control due to its simplicity 
and accuracy for piecewise constant inputs. 
The bilinear transformation (also known as Tustin transformation)  is valued for its frequency-warping properties, 
which preserve stability and provide a close approximation of continuous-time frequency responses.
Higher-order hold techniques, such as first-order hold (FOH) and second-order hold (SOH), 
reduce discretization errors by approximating input trajectories with higher-order polynomials.

The bilinear transformation and the Euler transformation are two types of commonly used methods 
due to their simplicity \textsuperscript{\cite{Kim19-NAD}}.
These two methods are formulated in equations \eqref{eq1} and \eqref{eq2} respectively:
\begin{equation} s = \frac{2}{T}\frac{z-1}{z+1}\label{eq1}\end{equation}
\begin{equation} s = \frac{z-1}{T}\label{eq2}\end{equation}
where T is the sampling period.
However, the frequency-warping phenomenon occurs near the Nyquist frequency. 

The Al-Alaoui integrator\textsuperscript{\cite{Alaoui08-AAO}} is formulated in equation \eqref{eq3}.
This method interpolates the trapezoidal and the rectangular integration rules.
\begin{equation} s = \frac{2}{T}\frac{z-1}{(1+a)z+(1-a)}\label{eq3}\end{equation}
where $a$ is a design parameter and $a \in [0,1]$.

The $\alpha$-approximation\textsuperscript{\cite{Sekara05-alpha}}, formulated in equation \eqref{eq_alpha}, 
is derived from the first-order approximation of both the numerator and the denominator.
\begin{equation} s = \frac{1}{T}\frac{z-1}{\alpha z+(1-\alpha)}\label{eq_alpha}\end{equation}
where $\alpha$ is a design parameter and $0\alpha \in [0,1]$.
Interestingly, \cite{Alaoui06-AAO} showed that the $\alpha$-approximation and the Al-Alaoui operator are the same.

In \cite{Kim19-NAD}, an accurate discretization method is presented, formulated as follows:
\begin{equation} s = \frac{1+\alpha_p}{T}\frac{z-1}{z+\alpha_p}\label{eq5}\end{equation}
where $\alpha_p$ is a design parameter and $\alpha_p \in [0,1]$.
However, this method is also equivalent to $\alpha$-approximation as explained in Section two of this paper.

In \cite{Zhang09-GBT}, a generalized bilinear transformation named GBT is studied.
This provides a class of digital approximations of an analog controller.
The transformation is illustrated in equation \eqref{eq_gbt}:
\begin{equation} s = \frac{1}{T}\frac{z-1}{\alpha_g z+(1-\alpha_g)}\label{eq_gbt}\end{equation}
where $\alpha_g$ is design parameter and belongs to the interval $(-\infty, \infty)$.
Compare \eqref{eq_gbt} with \eqref{eq_alpha}, 
it can be seen that the GBT is very similar to the $\alpha$-approximation despite the range of the design parameter $\alpha$.

This paper studies $\alpha$-approximation deeply which is a more generalized form of the first-order approximation.
It is divided into seven sections including an introduction a conclusion. 
Section two introduces the $\alpha$-approximation method and its relationship with other existing methods. 
The third section presents the mathematical interpretation of the $\alpha$-approximation.
The fourth section presents the stability analysis.
The fifth section presents the distortion analysis.
The sixth section presents the exclusion hypothesis, which posits that the amplitude distortion and phase distortion 
can't be minimized simultaneously by regulating parameter $\alpha$ alone.

\section{The $\alpha$-Approximation}
\subsection{Definition of $\alpha$-Approximation}
In the process of discretization of analog systems, the well-known mapping from the s-domain to the z-domain can be used by substitution.
\begin{equation} z = e^{sT}\label{eq_exact_disc}\end{equation}
This transformation maps the left half of the s-plane into the interior of the unit circle in the z-plane.

Starting from the basic transformation \eqref{eq_exact_disc}, the equivalent relation is defined as follows: 
\begin{equation} z = e^{sT} = e^{s[(1-\alpha)T+\alpha T]} = \frac{e^{s(1-\alpha)T}}{e^{s\alpha T}}, \alpha \in [0,1]\label{eq_alpha_appr}\end{equation}

Using the Taylor expansion for both the numerator and the denominator on the right side of expression \eqref{eq_alpha_appr} and neglecting all terms of second order and higher.
The expression \eqref{eq_alpha_appr} becomes:
\begin{equation} z = \frac{\sum_{n=0}^{\infty}\frac{[s(1-\alpha)T]^n}{n!}}{\sum_{m=0}^{\infty}(-1)^k \frac{(s\alpha T)^m}{m!}} \approx \frac{1+s(1-\alpha)T}{1-s\alpha T}\label{eq_alpha_appr2}\end{equation}

Solving equation \eqref{eq_alpha_appr2} for the variable $s$ yields the first-order approximation, which is defined as $\alpha$-approximation.
\begin{equation} s = \frac{1}{T}\frac{z-1}{1+\alpha(z-1)}\label{eq_alpha_appr3}\end{equation}
Where $\alpha$ is design parameter and belongs to the interval [0, 1].

\subsection{Mapping of $\alpha$-Approximation}
\begin{figure}[h] 
  \centering
  \includegraphics[width=\linewidth]{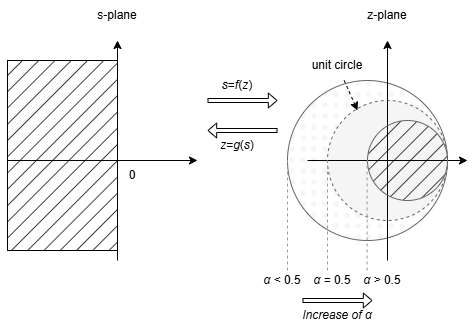}
  \caption{Mapping of s-plane to z-plane}
  \label{fig:mapping}
\end{figure}

Figure~\ref{fig:mapping} shows the mapping of the s-plane to the z-plane.
According to the theory of stability of discrete systems, the mapping is stable unless the left half of the s-plane is mapped into the unit circle of the z-plane. 
Figure~\ref{fig:mapping} shows clearly that the mapping is stable when $\alpha \geq 0.5$. 
However, the mapped z-plane exceeds the boundaries of the unit circle if $\alpha < 0.5$. 

\subsection{Relationship with Existing Method}
The relationship between $\alpha$-approximation and some existing methods is shown in Table~\ref{tab:relation}. 
The Euler and Tustin transformation are two specific forms of $\alpha$-approximation, with $\alpha$ equals 1 and 0.5, respectively.
The method in \cite{Kim19-NAD} is the same as the $\alpha$-approximation. 
The only difference between them is the parameters, and their relationship is expressed as follows:
\begin{equation} \alpha=\frac{1}{1+\alpha_p}\label{eq_relation_1}\end{equation}

The Al-Alaoui operator \textsuperscript{\cite{Alaoui08-AAO}} is equivalent to the $\alpha$-approximation. 
The difference is the design parameters, and their relationship is expressed as follows:
\begin{equation} \alpha=\frac{1+a}{2}\label{eq_relation_2}\end{equation}

In this case, $\alpha$ is limited to [0.5, 1] because the range of the parameter $a$ is [0, 1].
This result is also consistent with the nature of the Al-Alaoui operator, 
which interpolates the trapezoidal (Tustin) and the rectangular (Euler) integration rules.

The GBT \textsuperscript{\cite{Zhang09-GBT}} is also the same as the $\alpha$-approximation.
The design parameters are the same, and their relationship is expressed as follows:
\begin{equation} \alpha=\alpha_g\label{eq_relation_3}\end{equation}
The only difference is in the range of the design parameter, GBT extends the range of $\alpha$ from [0,1] to $(-\infty, \infty)$.

\begin{table*}[t] 
  \caption{Relations with existing methods}
  \label{tab:relation}
  \centering
  \begin{tabular}{llll}
    \toprule
    \textbf{Method} & \textbf{Transformation} & \textbf{Parameters} & \textbf{Relation}  \\
    \midrule
    $\alpha$-approximation        & $s = \frac{1}{T}\frac{z-1}{\alpha z+(1-\alpha)}$      & $\alpha \in [0,1]$               & \textbf{reference} \\
    Euler                         & $s = \frac{z-1}{T}$                                   & /                                & $\alpha$=1 \\
    Tustin                        & $s = \frac{2}{T}\frac{z-1}{z+1}$                      & /                                & $\alpha$=0.5\\
    Method in \cite{Kim19-NAD}    & $s = \frac{1+\alpha_p}{T}\frac{z-1}{z+\alpha_p}$      & $\alpha_p \in [0,1]$             & $\alpha=\frac{1}{1+\alpha_p}$\\
    Al-Alaoui \cite{Alaoui08-AAO} & $s = \frac{2}{T}\frac{z-1}{(1+a)z+(1-a)}$             & $a \in [0,1]$                    & $\alpha=\frac{1+a}{2}$\\
    GBT \cite{Zhang09-GBT}        & $s = \frac{1}{T}\frac{z-1}{\alpha_g z+(1-\alpha_g)}$  & $\alpha_g \in (-\infty, \infty)$ & $\alpha=\alpha_g$\\
    \bottomrule
  \end{tabular}
\end{table*}

\section{Mathematical Interpretation}

\begin{figure}[h]  
  \centering
  \includegraphics[width=\linewidth]{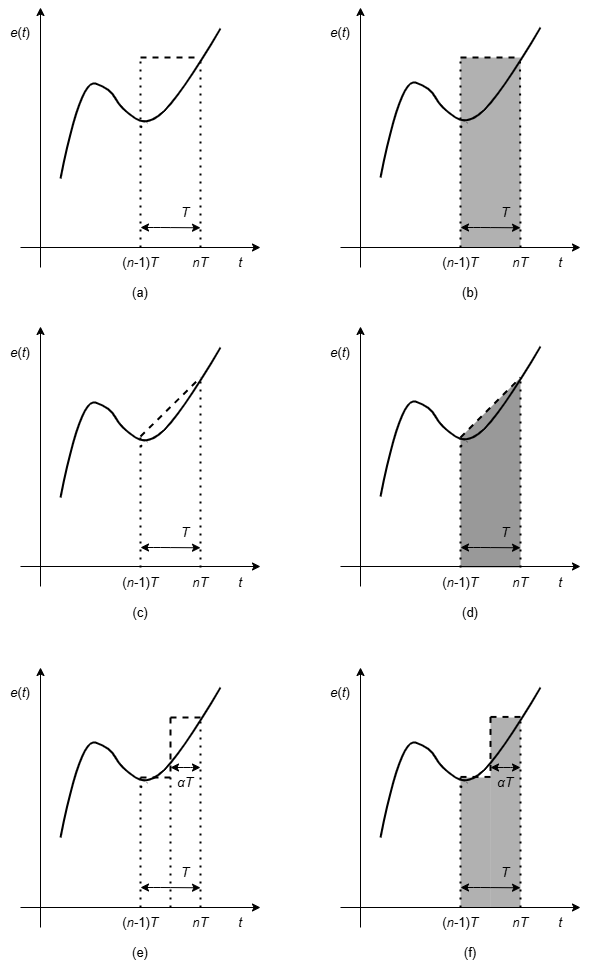}
  \caption{(a)-(b): Rectangular approximation (Euler), (c)-(d): Trapezoidal approximation (Tustin), (e)-(f): Hexagonal approximation ($\alpha$-approximation)}
  \label{fig:math_interpretation}
\end{figure}

Starting from the relationship between the error function, $e(t)$, and the original function, $u(t)$, as illustrated below:
\begin{equation} e(t) = \frac{d u(t)}{d t} \label{eq_diff}\end{equation}

In the continuous domain, $u(t)$ is expressed as follows: 
\begin{equation} u(t) = \int e(t) \,dt  \label{eq_u_e_cont}\end{equation}

In the discrete domain, $u(n)$ is expressed as follows:
\begin{equation} u(n) = \int_{(n-1)T}^{nT} e(t) \,dt + u(n-1) \label{eq_u_e_disc}\end{equation}

Figure~\ref{fig:math_interpretation} compare the mathematical interpretation of the $\alpha$-approximation and other methods.

Figure~\ref{fig:math_interpretation}(a) illustrates the Euler transformation (also known as the backward-difference approximation) geometrically.
The area enclosed by the three dotted lines and the horizontal axis is rectangular.
This rectangular approximation equals the numerical integration of the solid area as illustrated in Figure~\ref{fig:math_interpretation}(b).
In this case, the error function $e(t)$ is expressed as follows:
\begin{equation} e(t) = e(n) \label{eq_rect_error}\end{equation}
In the discrete domain, $u(n)$ is expressed as follows:
\begin{equation} u(n) = \int_{(n-1)T}^{nT} e(t) \,dt + u(n-1) = e(n)\cdot T + u(n-1) \label{eq_u_rect_appr}\end{equation}

Figure~\ref{fig:math_interpretation}(c) illustrates the Tustin transformation geometrically.
The area enclosed by the three dotted lines and the horizontal axis forms a trapezoid.
This trapezoidal approximation equals the numerical integration of the solid area as illustrated in Figure~\ref{fig:math_interpretation}(d).
In this case, the error function $e(t)$ is expressed as follows:
\begin{equation} 
  \begin{split}
      e(t) &= \frac{e(n)-e(n-1)}{T} \cdot [t-(n-1)T] + e(n-1) \\  
            & ,(n-1)T < t < n\cdot T      
  \end{split}
  \label{eq_trap_error}
\end{equation}
In the discrete domain, $u(n)$ is expressed as follows:
\begin{equation} u(n) = \frac{e(n)+e(n-1)}{2} \cdot T + u(n-1) \label{eq_u_trap_appr}\end{equation}

Figure~\ref{fig:math_interpretation}(e) illustrates the $\alpha$-approximation geometrically.
The area enclosed by the three dotted lines and the horizontal axis is hexagonal. 
This is why it is called the hexagonal approximation.
The hexagonal approximation equals the numerical integration of solid area as illustrated in Figure~\ref{fig:math_interpretation}(f).
In this case, error function $e(t)$ is expressed as follows:
\begin{equation} 
  e(t) = 
  \begin{cases}
      e(n-1), & t \in [(n-1)T, (n-\alpha)T] \\
      e(n), & t \in ((n-\alpha)T, n\cdot T]
  \end{cases}
  \label{eq_hexa_error}
\end{equation}
In the discrete domain, $u(n)$ is expressed as follows:
\begin{equation} u(n) = (1-\alpha)\cdot e(n-1)T + \alpha \cdot e(n)T + u(n-1) \label{eq_u_hexa_appr}\end{equation}
Therefore,
\begin{equation} (1 - z^{-1})\cdot U(z) = [(1-\alpha)\cdot z^{-1}\cdot T + \alpha\cdot T] \cdot E(z)\label{eq_Uz_Ez}\end{equation}
\begin{equation} s = \frac{E(z)}{U(z)} = \frac{1}{T}\frac{1-z^{-1}}{\alpha + (1-\alpha) \cdot z^{-1}}\label{eq_s2z_hexa}\end{equation}

Moreover, the hexagonal area in Figure~\ref{fig:math_interpretation}(f) comprises two rectangular parts.
The left part is a forward rectangular area and the right part is a backward rectangular area.
The physical meaning of the parameter $\alpha$ is the percentage of the backward rectangular area, 
while $(1-\alpha)$ is the percentage of the forward rectangular area.

\begin{figure*}[h]  
  \centering
  \includegraphics[width=0.7\linewidth]{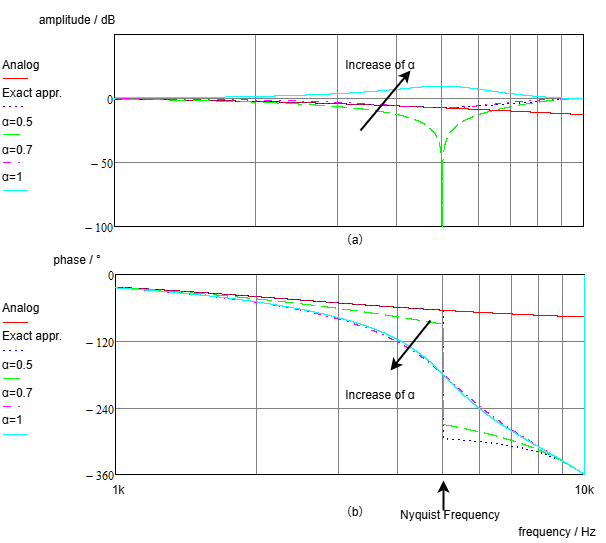}
  \caption{Amplitude and phase distortion with different $\alpha$}
  \label{fig:distortion}
\end{figure*}

\section{Stability Analysis}
Let $s=\sigma_s + j\omega_s$ and $z=Re(z) + j\cdot Im(z)$, we have the following expression by substituting these into equation \eqref{eq_alpha}:
\begin{equation}   
  \begin{split}
      s &= \sigma_s + j\omega_s = \frac{1}{T}\frac{[Re(z) + j\cdot Im(z)]-1}{\alpha (Re(z) + j\cdot Im(z))+(1-\alpha)} \\  
         &=\frac{1}{T}\frac{[(Re(z)-1)+j\cdot Im(z)][(\alpha Re(z)+1-\alpha)-j\alpha\cdot Im(z)]}{[\alpha Re(z)+1-\alpha]^2+[\alpha\cdot Im(z)]^2}  \\    
         &=\frac{1}{T}\frac{[\alpha(Re(z)-1)^2+Re(z)-1+\alpha\cdot Im(z)^2]+j\alpha\cdot Im(z)}{[\alpha Re(z)+1-\alpha]^2+[\alpha\cdot Im(z)]^2}    \\    
  \end{split}
  \label{eq_stab1}
\end{equation}
Therefore, $\sigma_s$ and $\omega_s$ can be derived as follows:
\begin{equation} \sigma_s = \frac{1}{T}\frac{\alpha(Re(z)-1)^2+Re(z)-1+\alpha\cdot Im(z)^2}{[\alpha Re(z)+1-\alpha]^2+[\alpha\cdot Im(z)]^2} \label{eq_sigma_s}\end{equation}
\begin{equation} \omega_s = \frac{1}{T}\frac{Im(z)}{[\alpha Re(z)+1-\alpha]^2+[\alpha\cdot Im(z)]^2} \label{eq_omega_s}\end{equation}
As previously mentioned, the transformation is stable unless the left half of the s-plane is mapped into the unit circle of the z-plane, which implies that $\sigma_s \leq  0$.
Substituting this into equation \eqref{eq_sigma_s} yields,
\begin{equation} \alpha(Re(z)-1)^2+Re(z)-1+\alpha\cdot Im(z)^2 \leq 0\label{eq_stab2}\end{equation}
Therefore, we have:
\begin{equation} [Re(z) - (1-\frac{1}{2\alpha})]^2 + Im(z)^2 \leq (\frac{1}{2\alpha})^2\label{eq_stab3}\end{equation}
This equation has two crossing points on the real axis, labeled as $\sigma_{z1}$ and $\sigma_{z2}$,
\begin{equation} \sigma_{z1} = 1, \sigma_{z2} = -\frac{1}{2\alpha}\label{eq_stab4}\end{equation}
Since $z$ should be within the unit circle of the z-plane, therefore, we have:
\begin{equation} \sigma_{z2} = -\frac{1}{2\alpha} \geq -1\label{eq_stab5}\end{equation} 
Therefore, the restriction for a stable transformation can be expressed as follows:
\begin{equation} \alpha \geq 0.5 \label{eq_stab6}\end{equation}

In conclusion, the parameter $\alpha$ should be within the range [0.5, 1] in order to achieve a stable transformation, 
Specifically, when $\alpha$ is set to 0.5 or 1, the $\alpha$-approximation turns into the Tustin transformation or Euler transformation, respectively.

\section{Distortion Analysis}

Since the $\alpha$-approximation is a first-order approximation, it will inevitably introduce distortion including amplitude and phase errors.
Figure~\ref{fig:distortion} shows the amplitude and phase distortion of a low-pass filter (LPF) with different $\alpha$.
The sampling frequency is 10kHz, therefore the Nyquist frequency is 5kHz.
The LPF's crossing frequency is 2.4kHz.

Figure~\ref{fig:distortion}(a) shows that amplitude distortion can be reduced to minimum when $\alpha$ is around 0.7.
However, the phase distortion can be reduced to minimum when $\alpha$ is 0.5, as illustrated in figure~\ref{fig:distortion}(b).

\section{Exclusion Hypothesis}
Based on our research into several cases, including low-pass filter, the proportional-integral controller, 
the proportional-resonant controller, and the notch filter, 
we found that it was impossible to minimize the amplitude distortion and phase distortion simultaneously by regulating parameter $\alpha$ alone.
For this reason, we proposed the following hypothesis: 

\textbf{Exclusion Hypothesis:} Suppose the analog system is internally stable. 
Then, for every frequency $f$ in the range $f(0, f_{nyquist})$, 
there is no single parameter $\alpha$ to minimize the amplitude distortion and the phase distortion simultaneously.

Machine learning models, adept at identifying subtle patterns in vast datasets, 
are being used to function as "conjecture generators".  
These models propose mathematical statements that appear to be true based on available data\textsuperscript{\cite{Raayoni2021},\cite{DeLaVina2019}}.
Perhaps the most significant impact of machine learning has been in assisting the process of formal proof, 
either by guiding human mathematicians or by steering automated theorem provers\textsuperscript{\cite{Lample2022}}.
We are currently working on verifying this hypothesis using machine learning, and will provide an update once results are available.

\section{Conclusion}
We conducted a thorough analysis of the $\alpha$-approximation method, 
including its mathematical interpretation, stability analysis and distortion analysis.
The main contributions of this work can be summarized as follows:
\begin{itemize}
    \item The $\alpha$-approximation can be derived by numerically integrating the error function and is defined as hexagonal approximation.
    \item We propose a physical interpretation of parameter $\alpha$ as the percentage of the backward rectangular area.
    \item The stable range of the parameter $\alpha$ is rigorously established as [0.5, 1] through a detailed mathematical proof based on stability analysis.
    \item We demonstrate that amplitude distortion and phase distortion are significant when the frequency approaches the Nyquist frequency as shown in the Bode plot.
    \item We propose the exclusion hypothesis, which posits that there is no single parameter $\alpha$ to minimize the amplitude distortion and phase distortion 
    simultaneously across all frequency points within the Nyquist frequency range.
\end{itemize}


\bibliographystyle{IEEEtran}

\bibliography{reference} 

\end{document}